\documentclass[useAMS,usenatbib,usegraphicx,onecolumn]{mn2e}

\usepackage{amssymb,amsfonts,amsmath,amstext,amsgen,amsopn,amsxtra,indentfirst,lscape,times,rotating}

\title[HD 69830 debris disc mass]{Estimating the mass of the debris disc in HD 69830}
\author[Heng]{Kevin Heng$^{1}$\thanks{E-mail: kheng@phys.ethz.ch (KH)}\\
$^{1}$Zwicky Fellow, ETH Z\"{u}rich, Institute for Astronomy, Wolfgang-Pauli-Strasse 27, CH-8093, Z\"{u}rich, Switzerland}

\begin{document}

\date{Submitted 2011 March 22.  Re-submitted 2011 April 19.  Accepted 2011 April 21.}

\pagerange{\pageref{firstpage}--\pageref{lastpage}} \pubyear{2010}

\maketitle

\label{firstpage}

\begin{abstract}
We present a method to estimate the mass of the debris disc in the HD 69830 system, which also hosts three exoplanets with Neptune-like minimum masses.  By considering the range of published stellar ages, we interpret the infrared emission from the debris disc as originating from a steady state, collisional cascade of dust grains.  Using dynamical survival models subjected to observational constraints, we estimate the allowed range of disc masses.  If the disc has an age $t_\star \approx 1$ Gyr, then its mass is $M_{\rm disc} \approx 3$--$4 \times 10^{-3} M_\oplus$, several times more massive than our asteroid belt.  The maximum allowed age for the disc and the number of planetesimals it contains are determined by the assumed value for the binding energy of the planetesimals.  If one insists on interpreting the disc as being transient, then this mass estimate becomes an upper limit.
\end{abstract}

\begin{keywords}
minor planets, asteroids -- planets and satellites: general
\end{keywords}

\section{Introduction}
\label{sect:intro}

Debris discs are believed to arise from planetesimals colliding and producing a cascade of dust grains, which reprocess the starlight into infrared emission \citep{z01,wyatt08}.  While infrared and sub-mm observations yield reasonable estimates for the dust mass, estimating the \emph{disc mass} is generally challenging because the planetesimals cannot be detected via conventional methods \citep{ht10}.

Of the hundreds of debris discs studied, the one residing in HD 69830 stands out because the system also hosts three exoplanets with Neptune-like minimum masses \citep{lovis06}.  Steady state models using a narrow interpretation of the disc properties infer that the infrared emission is anomalously bright, which implies that we may be observing HD 69830 at a special time in its evolution --- it is a ``transient" debris disc fed by a recent, violent collision between planetesimals \citep{wyatt07,ht10}.  \cite{lisse07} reach a more subdued conclusion: based on the discovery of icy dust grains within the debris disc, they remark that the dust was either produced within the last year (relative to the time their observations were taken) or was being continuously replenished by planetesimal collisions.

In this study, we offer an alternative, plausible perspective: as the stellar/disc age (assuming they are the same) is uncertain, it is possible to still interpret the disc emission as being non-transient.  In this case, the old age of the disc becomes a friend and not a foe, because the possible values for the disc mass become severely restricted by both dynamical and observational constraints.  In particular, we show that the disc in HD 69830 is several times more massive than the asteroid belt in our Solar System.  In \S\ref{sect:obs}, we review the observational constraints on the HD 69830 system.  In \S\ref{sect:models}, we apply dynamical survival models to the system.  We conclude by discussing the implications of our study in \S\ref{sect:discussion}, including the main model uncertainties.

\section{Observational constraints}
\label{sect:obs}

The key aim is to meaningfully restrict the parameter space of our models using observational constraints.  Therefore, we begin by reviewing the existing observational constraints found in the literature.  Located 12.6 pc away, the HD 69830 star has a spectral type K0V, an effective temperature $T_\star = 5385 \pm 20$ K and a luminosity $L_\star = 0.60 \pm 0.03 L_\odot$ \citep{lovis06}.  By using theoretical evolutionary tracks, \cite{lovis06} derived a stellar mass of $M_\star = 0.86 \pm 0.03 M_\odot$.

High precision radial velocity measurements using the HARPS spectrograph placed the exoplanets HD 69830b--d at semi-major axes of $a_p = 0.0785$, 0.186 and 0.630, respectively \citep{lovis06}.  The corresponding minimum masses are $M_p \sin i = 10.2$, 11.8 and 18.1 Earth masses, where $i$ denotes the inclination of the system (assuming co-planar orbits).

Infrared excesses at 12, 23.7 and 25 $\mu$m were detected in HD 69830; the optical depth (covering fraction) of the associated dust grains was inferred to be $\tau = 3 \times 10^{-4}$, more than 3 orders of magnitude higher than for the asteroid belt \citep{beich05}.  The non-detection of an infrared excess at 70 $\mu$m implies that the dust grains predominantly have sizes smaller than 70 $\mu$m$/2\pi \approx 11 ~\mu$m.  The strong mineralogical features detected in the spectrum from 8--35 $\mu$m further argues for small grain sizes \citep{beich05,lisse07}, which rules out the possibility of the infrared excess being due to a dynamically warm disc (with particle sizes $\gg 1 ~\mu$m; \citealt{ht10}).  Detailed modelling of the infrared spectral energy distribution yielded semi-major axes of $a=0.92$--1.16 AU \citep{lisse07}.  Constraints from mid-infrared interferometry place an upper limit of 2.4 AU on the disc size \citep{smith09}.  These estimates for $a$ are consistent with stability analyses which suggest that planetesimals may exist at $a=0.8$--1.2 AU \citep{lovis06,ji07}.  We thus adopt a conservative range of semi-major axes for the disc of $a=0.8$--1.3 AU.\footnote{Such a disc extends over one octave in semi-major axis, since it is centered at 1 AU and extends over $\Delta a=0.5$ AU.}

The most uncertain parameter in the system is the stellar age $t_\star$.  Ages of $t_\star \approx 0.6$--5 Gyr have been claimed (see \citealt{beich05} and references therein).  \cite{wyatt07} favour $t_\star=2$ Gyr (see their Table 1); in tandem with assuming a single value of $a=1$ AU, as well as fixed values for the parameters describing the planetesimals, they concluded that the infrared emission due to dust exceeds its steady state value, thereby qualifying it as a ``transient" (see also \citealt{ht10}).

\section{Method}
\label{sect:models}

\begin{figure}
\begin{center}
\includegraphics[width=0.6\columnwidth]{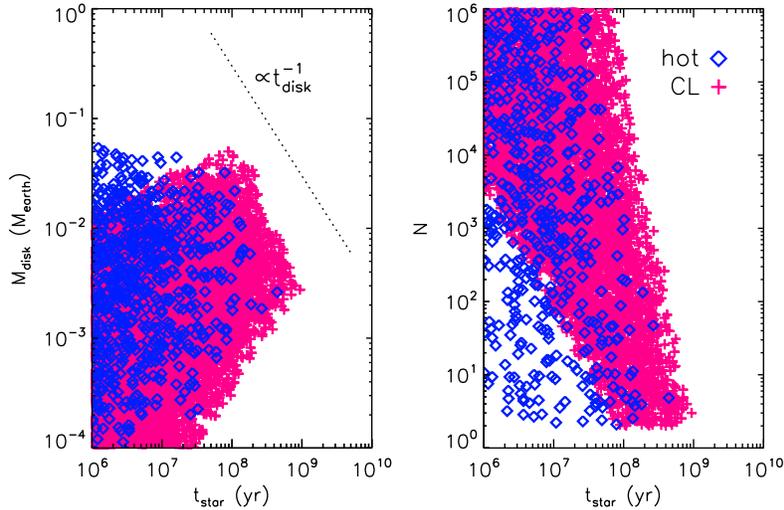}
\end{center}
\vspace{-0.2in}
\caption{Left panel: dynamical survival models exploring the allowed range of disc masses as a function of the stellar age.  Right panel: the corresponding allowed number of planetesimals.  Both dynamically hot and collision-limited (denoted by ``CL") discs are considered, with the latter being a special case of the former (see text).}
\label{fig:survival}
\end{figure}

The formation of planetary systems from first principles is a challenging task fraught with theoretical uncertainties --- these challenges are present for any formation model of the HD 69830 system \citep{alibert06}.  Since our main interest is in estimating the disc mass, a simpler approach is to construct a \emph{survival} model with the reasoning that any observed disc must have survived all dynamical processes, including gravitational instability, dynamical chaos, gravitational scattering, physical collisions and radiation forces that would lead to significant evolution over its lifetime \citep{ht10}.  From a dynamical viewpoint, discs may be classified as being dynamically hot, warm or cold.  Only dynamically hot discs are capable of producing a collisional cascade of dust grains.  Collision-limited discs are a special sub-class of dynamically hot discs, where the collisional time scale is equal to the disc age.  Dynamical survival models are relatively simple and may be described mainly by five parameters: the disc mass $M_{\rm disc}$, the disc semi-major axis $a$, planetesimal size $r$, planetesimal radial velocity dispersion $\sigma_r$ and the stellar/disc age $t_\star$.  The radial velocity dispersion $\sigma_r$ is in turn directly related to the root mean square eccentricity of the planetesimal orbits \citep{ht10}.

Since observational constraints exist for $a$ and $t_\star$ (see \S\ref{sect:obs}), we may thus use the models to constrain the allowed parameter space for the other three parameters.  (The stellar mass $M_\star$, luminosity $L_\star$ and radius $R_\star$ enter as minor parameters and are also fixed by the observations.)  Additionally, an important constraint is provided by the observed dust optical depth $\tau$ (see \S\ref{sect:obs}).  We permit only disc models with a theoretically calculated dust optical depth $\tau_0$ which satisfies the condition,
\begin{equation}
\epsilon^{-1} \tau_0 \le \tau \le \epsilon \tau_0.
\label{eq:tau}
\end{equation}
The dimensionless quantity $\epsilon$ allows for the uncertainties associated with our survival models, since the computed value of $\tau_0$ depends on $M_{\rm disc}$, $r$ and $\sigma_r$.  Smaller values of $\epsilon$ lead to narrower ranges of estimates for the disc mass.  We choose to be conservative and adopt $\epsilon = \sqrt{10}$ such that the computed dust optical depth varies by a factor $\epsilon^2 = 10$.

Figure \ref{fig:survival} shows the range of possible disc masses $M_{\rm disc}$ and the corresponding number of planetesimals in the disc $N$ as functions of $t_\star$.  These diagrams were constructed by first considering a wide range of values for $M_{\rm disc}$, $r$ and $\sigma_r$ via a Monte Carlo approach.  For each randomly generated trio of parameters, a list of dynamical conditions were checked, as described in detail in \cite{ht10}.  The condition in equation (\ref{eq:tau}) was also checked.  Only the discs which satisfy all of these conditions are retained.

\begin{figure}
\begin{center}
\includegraphics[width=0.52\columnwidth]{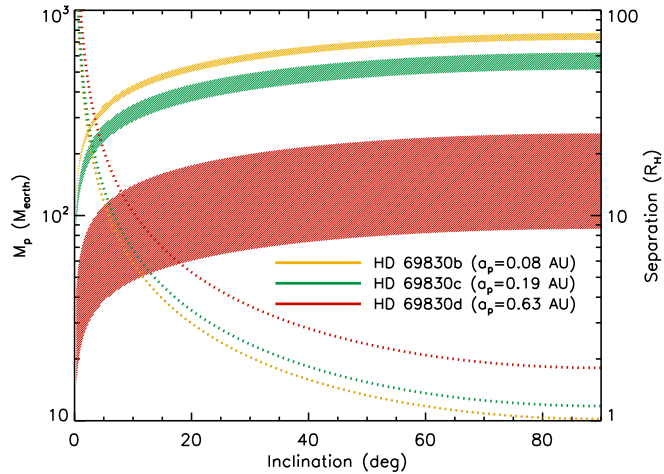}
\end{center}
\vspace{-0.2in}
\caption{Masses of the exoplanets HD 69830b--d (dotted lines) as well as the spatial separation between them and the debris disc (in terms of Hill radii $R_{\rm H}$; shaded regions), both as a function of the inclination angle $i$.}
\label{fig:schematic}
\end{figure}

A few features of Figure \ref{fig:survival} are worth pointing out.  Firstly, $M_{\rm disc}$ and $N$ are essentially unconstrained if the disc is allowed to be much younger than the star (e.g., as suggested by \citealt{payne09}).  For young disc ages, the enormous allowed range of values for $N$ reflects the fact that a disc of mass $M_{\rm disc}$ may be configured into a large variety of different planetesimal masses which still satisfy both the dynamical and observational constraints.  In other words, the disc mass and observed infrared emission from dust grains are degenerate quantities.  Secondly, if we allow the stellar/disc age to be $t_\star \approx 1$ Gyr, then decent estimates for the disc mass and number of planetesimals are produced:
\begin{equation}
\begin{split}
M_{\rm disc} &\approx 3 \mbox{--} 4 \times 10^{-3} M_\oplus,\\
N &\approx 2 \mbox{--} 3.
\end{split}
\label{eq:estimates}
\end{equation}
Curiously, \cite{beich05} extrapolated the inferred dust distribution to $\sim 10$--100 km-sized planetesimals and inferred a disc mass $\sim 10^{-3} M_\oplus$.  \emph{It is precisely because the disc is old and bright that the range of possibilities for its mass is limited.}

It is important to note that the approach of \cite{ht10} does not generally require $N \gg 1$ for dynamically hot discs.  For example, \S3.2.1 of that paper describes a generalization of the criterion for the Toomre stability of a disc (which does assume a fluid treatment) to discs where $N \sim 1$.  In other words, while the dust grains are assumed to be in a collisional cascade, the planetesimals are not.

It is reasonable to ask if any of the exoplanets in the HD 69830 system are capable of dynamically stirring the disc and thereby modifying the predicted dust content away from its steady state value.  Mutual gravitational interactions occur when two bodies are separated by less than about $3.5 R_{\rm H}$ \citep{chambers96}, where the mutual Hill radius is defined as \citep{ji07}
\begin{equation}
R_{\rm H} \approx \frac{a+a_p}{2} \left( \frac{M_p}{3M_\star} \right)^{1/3}.
\end{equation}
Figure \ref{fig:schematic} shows the spatial separations between HD 69830b--d and the disc, which we assume to extend from $a=0.8$ AU to 1.3 AU.  It is apparent that HD 69830b and HD 69830c are located at separations greatly exceeding $10 R_{\rm H}$ for non-zero values of the inclination.  Only for inclinations of $i \lesssim 4^\circ$ is HD 69830d capable of dynamically stirring the disc, which corresponds to $M_p \gtrsim 0.8 M_{\rm J}$ where $M_{\rm J}$ is the mass of Jupiter.  These values of $i$ are not ruled out by stability analyses of the planetary orbits \citep{lovis06,ji07}.  In such cases, the presence of HD 69830d may increase the radial velocity dispersion $\sigma_r$ of the planetesimals in the disc and therefore increase the eccentricities of their orbits \citep{raf03}, implying that collisions between them will become more violent.  Non-intuitively, however, the collision \emph{rate} either remains unchanged or \emph{decreases}, since \citep{ht10}
\begin{equation}
t^{-1}_c \propto
\begin{cases}
\mbox{constant}, & \Theta \ll 1,\\
\sigma^{-2}_r, & \Theta \gg 1,\\
\end{cases}
\end{equation}
depending on whether the self-gravity of the planetesimals is important (which is the case when the Safronov number $\Theta$ exceeds unity).  For the values of $M_{\rm disc}$ and $N$ inferred from Figure \ref{fig:survival}, each of the planetesimals are $\sim 1$--10 times the mass of Ceres, with radii $\sim 1000$ km, and most likely have $\Theta > 1$.

Several mean motion resonances of the planetesimals with HD 69830d have already been identified by \cite{alibert06}.  Specifically, the 2:1 resonance at $a \approx 1$ AU creates a zone of accumulation which allows the debris disc to survive.  It is expected that the mean motion resonances will not substantially modify the dust production properties of the planetesimals away from the steady state model we have used in this study.

We conclude that our dynamical survival models give reasonable estimates for the disc mass if $i > 4^\circ$.

\section{Discussion}
\label{sect:discussion}

Using the existing observational constraints on the HD 69830 system, we have applied the dynamical survival models of \cite{ht10} to estimate the mass of the debris disc and find it to be several times more massive than the asteroid belt in our Solar System ($\approx 6 \times 10^{-4} M_\oplus$; \citealt{kra02}).  Our estimate hinges on being able to interpret the HD 69830 debris disc within the context of a steady state model, which is only possible if one adopts the lower range of stellar ages quoted for HD 69830 ($t_\star \approx 1$ Gyr).  If one insists on interpreting the disc as being transient, then this mass estimate becomes an upper limit.

The most uncertain piece of physics involved in the dynamical survival models concerns the collisions between the parent planetesimals responsible for dust production.  The binding energy per unit mass of a planetesimal is typically described by the function \citep{ba99},
\begin{equation}
Q^\ast = Q_0 \left[ \left( \frac{m}{m_0} \right)^a + \left( \frac{m}{m_0} \right)^b \right],
\end{equation}
where the quantities $Q_0$, $m_0$, $a$ and $b$ are parameters determined from a combination of simulations and laboratory experiments.  The specification of the values for the indices $a$ and $b$ also allows the mass distribution of the planetesimals to be fixed \citep{ht10}.  Following \cite{ht10}, we have adopted $m_0=10^{14}$ g, $Q_0=6 \times 10^5$ erg g$^{-1}$, $a=-0.13$ and $b=0.44$ by taking an average of the values assumed for ice and basalt, while being aware that they are somewhat uncertain.  

In particular, the value of $Q_0$ may vary by several orders of magnitude \citep{sl09}.  For example, decreasing $Q_0$ to $6 \times 10^4$ erg g$^{-1}$ results in similar estimates for $M_{\rm disc}$ (equation [\ref{eq:estimates}]), but only allows for discs with $t_\star \lesssim 0.7$ Gyr to survive.  The relative ease of breaking apart the planetesimals means that they are less likely to survive for a given time.  If $Q_0 = 6 \times 10^3$ erg g$^{-1}$, then we now have $t_\star \lesssim 0.2$ Gyr, which is inconsistent with the lowest value of the stellar age quoted in the literature (0.6 Gyr; see \S\ref{sect:obs}).  Also, as $Q_0$ decreases, the mass $M_{\rm disc}$ is distributed among a larger number $N$ of planetesimals.  For comparison, we note that \cite{wyatt07} adopt a constant $Q^\ast = 2 \times 10^6$ erg g$^{-1}$.  Our estimates highlight the fact that whether the disc is transient depends on the assumed value for the binding energy of the planetesimals.  This quantity in turn depends on the material composition of the planetesimals, for which we only have Solar System asteroids and Kuiper Belt objects as a guide.

\vspace{0.2in}
\noindent
\textit{K.H. acknowledges generous support from the Zwicky Prize Fellowship and the Star and Planet Formation Group at ETH Z\"{u}rich.  He is grateful to Michael Meyer and Vincent Geers for useful conversations, and thanks Christine Chen for encouraging him to publish this work.}


\label{lastpage}

\end{document}